\documentclass[aps,reprint,amsmath,amssymb]{revtex4-2}
\usepackage{graphics}
\usepackage{svg}
\usepackage{subfigure}
\usepackage{epsfig}
\usepackage{epsf,epic}
\usepackage{dcolumn}
\usepackage{color}
\usepackage{hyperref}
\usepackage{marvosym}
\usepackage{subfigure}
\usepackage{amsmath}
\usepackage{amssymb}
\usepackage{amsfonts}
\usepackage{wrapfig}
\usepackage{pstricks}
\usepackage{multirow}
\usepackage{float}
\usepackage{placeins}
\usepackage{bm}
\usepackage{threeparttable}%
\graphicspath{{./figs/}}
\usepackage{dcolumn}

\begin{document}
\large
\title{Phonon, Infrared  and Raman Spectra of LiGa$_5$O$_8$ from Density Functional Perturbation Theory}

\author{Sarker Md. Sadman}
\author{Walter R. L. Lambrecht}\email{walter.lambrecht@case.edu}

\affiliation{Department of Physics, Case Western Reserve University, 10900 Euclid Avenue, Cleveland, Ohio 44106-7079, USA}

\begin{abstract}
\large
LiGa$_5$O$_8$ with a cubic spinel type structure was recently reported to be a ultra-wide-band-gap semiconductor with unintentional p-type conduction. While the origin of p-type doping is still unclear, the fundamental properties of this material are of interest. Here we present a first-principles study of the phonons using density functional perturbation theory. The phonon band structures show no unstable modes verifying the stability of the structure. We focus mainly on the phonons at the Brillouin zone center for which a full symmetry analysis is presented. We present the dielectric function contributions from the infrared active modes as well as the Raman spectra and their polarization dependence. The phonon density of states integrated over the Brillouin zone is also presented as this may related to disordered Raman spectra.
 \end{abstract}

\maketitle
\section{INTRODUCTION}
\large
LiGa$_5$O$_8$ was recently reported to be a p-type ultra-wide-band-gap (UWBG) semiconductor with a gap of 5.34 eV and cubic spinel structure \cite{Zhao2023}. This is an unusual finding as native p-type doping in UWBG oxides is rare and p-type doping is often hindered by small polaron formation. The experimental gap \cite{Zhao2023} extracted from a Tauc plot is slightly smaller than the calculated gap of 5.72 eV \cite{Lambrecht2024} using the quasiparticle self-consistent (QS)$G\hat{W}$  method, which is a many-body perturbation theory (MBPT)  method in which the self-energy $\Sigma=iGW$ is calculated from the one particle Green's function $G$ and the screened Coulomb interaction $\hat{W}$ including electron-hole interactions in the screening. The optical band gap obtained from the Bethe Salpeter Equation (BSE) in this same work \cite{Lambrecht2024} indicates a large exciton binding energy with lowest direct exciton gap of 5.5 eV and also showed the existence of a 0.1 eV lower indirect gap. The band gap is thus fairly well established. However, the origin of the p-type doping is not. Native defect calculations \cite{Dabsamut2024} showed all plausible acceptors to have high binding energies exceeding 0.75 eV. Furthermore, compensating shallow donors like Ga$_{\rm Li}$  antisites or deep donor oxygen vacancies were found to pin the equilibrium Fermi level deep in the gap thus predicting insulating behavior. Similar findings were found with evidence of small polaron formation for native defects and some candidate acceptor impurities by Lyons \cite{Lyons2024}. Insulating behavior has meanwhile been found upon annealing in O$_2$ and the stoichiometry conditions under which p-type behavior occurs have been more clearly established \cite{Kaitian25} to be somewhat O-poor, Li-poor and Ga-rich. In spite of the lack of understanding of the origin of p-type doping rectification behavior at pn-heterojunçtions between n-type Ga$_2$O$_3$ and p-type LiGa$_5$O$_8$ has been reported \cite{Vangipuram25}. LiGa$_5$O$_8$ has previously received some interest as a host for phosphorescent dopants like Cr \cite{DeClercq17,Sousa20,Huang2018}. The spinel structure of this compound was reported by Joubert {\sl et al.}\cite{Joubert63}. While the present paper does not address directly the question of p-type doping, the renewed interest in this material motivates  a broader study of its fundamental properties. Here we present a first-principles study of the phonons with the aim to assist on-going characterization efforts on this new material. First, the phonon band structure calculations are important to verify the mechanical stability of the material in this structure. Second,  Raman spectra are often used to characterize the crystalline quality of a material. Third, electron-phonon coupling is responsible for finite temperature gap renormalization and for  limitations to the mobility.   We here focus on the phonons at the Brillouin zone center and present a symmetry analysis of the vibrational modes of the 56 atom unit cell and report the calculated Infrared and Raman spectra as well as related Born effective charges and static and high frequency dielectric constants. 
In prior work, we studied the phonons and related properties in LiGaO$_2$, \cite{Boonchun2010} another compound in the Li-Ga-O system but that compound has a simpler 16 atom wurtzite based structure. Our results on that material obtained with a similar approach as here were in good agreement with experimental data. 

\section{COMPUTATIONAL METHOD} \label{sec:method}
 \large
 All the calculation presented here were carried out using the Projector Augmented Wave (PAW) method \cite{PAW} as implemented in the  {\sc Abinit} code \cite{Gonze2020,Torrent2008}. The following electron configurations were adapted as valence electrons: $1s^22s^1$ for Li, $4s^24p^13d^{10}$ for Ga, and $2s^22p^4$ for O. We used the generalized gradient approximation (GGA) in the PBESol parameterization \cite{PBEsol} as exchange correlation functional. We start from the structures provided by the Materials Project \cite{MAtProj,liga5o8struc,icsdliga5o8} and optimize the ground state structure first before proceeding with the electronic band structure calculation and subsequently studying the vibrational properties using density functional perturbation theory (DFPT) \cite{Gonze1997,Gonze1997a}. 
 DFPT relies on the $2n+1$ theorem which allows one to calculate up to third order derivatives of the total energy using first-order corrected wave functions. This allows one to obtain second derivatives versus atomic positions (force constants), second and third derivatives with respect to long range electric fields (high-frequency dielectric constants and second order susceptibilities in the static but electronic screening only limit), mixed second derivatives versus electric field and atomic displacements (Born effective charges as needed for the LO-TO splitting). For the Raman intensities one needs the derivative of the electric susceptibility versus atomic displacements, which is a mixed third order derivative\cite{Veithen2005a}.
Within the PAW or ultrasoft pseudopotential methods, however, the calculation of the Raman tensor requires second order corrected wave functions \cite{Miwa2011}
 
 The wave function cut-off used is 60 Ha and the PAW energy cut-off for the double grid was chosen to be 120 Ha. The structural relaxation was carried out with the tolerance for forces less than $5.0\times10^{-5}$Ha/Bohr. For the Brillouin zone sampling of the elecronic properties, an unshifted Monkhorst-Pack grid of size $8\times8\times8$ was used.
 To obtain phonon dispersion and an accurate density of phonon states (DOS), we use the interpolation 
 approach of \cite{Gonze1997a} using  analytic expressions for the long-range  force  constants and real space short-range force constants extracted from an initial coarse $2\times2\times2$ {\bf q}-mesh by Fourier transformation. The final {\bf q}-mesh used for the phonon DOS was $4\times4\times4$ with an energy cut-off of 25 Ha.

 For, the calculation of microscopic dielectric properties, IR, and Raman spectra, a zone centered phonon calculation was performed. The convergence criterion on forces used for the phonon calculations is $1.0\times10^{-7}$Ha/Bohr but for the higher order corrections in wavefunction the convergence criterion was chosen in terms of wavefunction residual of $1.0\times10^{-14}$.
\section{RESULTS AND DISCUSSION}
\large
\subsection{Crystal Structure}
\begin{figure}
\includegraphics[width=9cm]{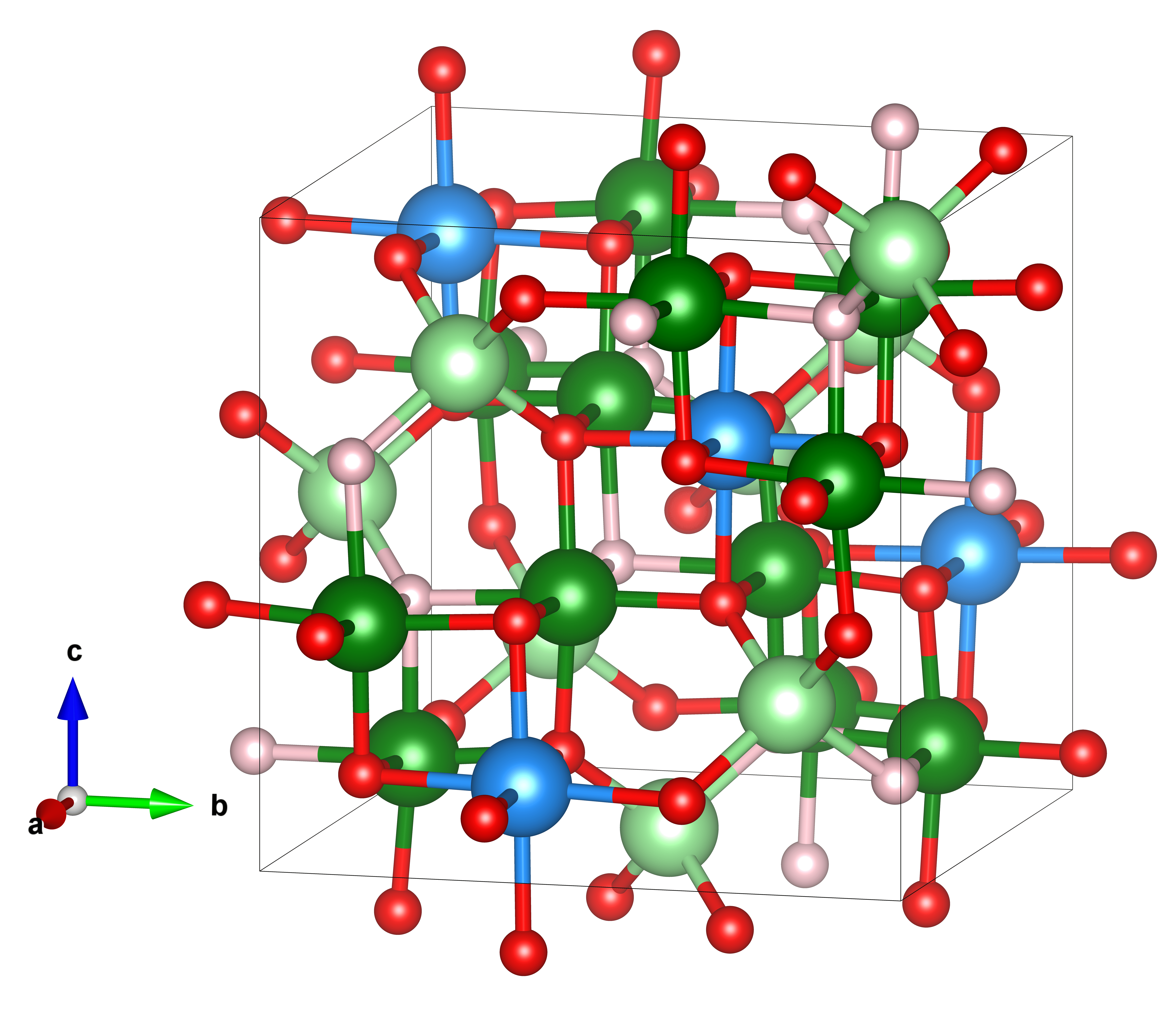}
  \caption{\large Crystal Structure of LiGa$_5$O$_8$, with 4 Li sites(blue), 8 tetrahedral Ga sites(light green), 12 octahedral Ga sites(dark green) and 8 $O^{(1)}$ oxygen sites in pink and 24 $O^{(2)}$ oxygen sites are in red.\label{crystal}}
\end{figure}
Figure~\ref{crystal} represents the spinel-type LiGa$_5$O$_8$ structure which contains four formula units. The  unit cell contains four Li  in the Wyckoff 4b positions, which are octahedrally coordinated, eight tetrahedral Ga (8c) sites, twelve octahedral Ga (12d) sites, and thirty-two O in two distinct Wyckoff sites O$^{(1)}$ in 8c positions  and O$^{(2)}$ in 24e positions. In total, there are  56 atoms in the  unit cell. The structure is cubic with space group $P4_33\overline{2}$ (No.212) and  point group $O$ in the Schoenflies notation or $432$ in the International Crystallography notation.

The calculated optimized parameters are compared with the experimental ones and previous calculations   in Table~\ref{lattice}.
Excellent agreement is obtained with experiment for the cubic lattice constant with a slight 0.3 \% overestimate and the reduced coordinates  of the the Wyckoff positions agree well with the Material Project optimization to at least 3 digits.  
The bond lengths are reported in Table~\ref{lattice}.
\begin{table*}
\large
  \caption{Structural parameters of LiGa$_5$O$_8$. \label{lattice}}
  \begin{ruledtabular}
    \begin{tabular}{lcc}
      Parameters & Theory & Expt.\\
      \hline
      Lattice constant a (\AA)		& 8.233 & 8.203\cite{Joubert63} \\ \hline
      \multicolumn{3}{c}{Wyckoff positions} \\\hline
      4d Li & (5/8, 5/8, 5/8) & (5/8, 5/8, 5/8) \\
      8c Ga & (0.253845, 0.253845, 0.253845)              & (0.253299, 0.253299, 0.253299)\cite{liga5o8struc} \\
      12d Ga & (1/8, 0.365375, 0.884623) & (1/8, 0.365985, 0.884015) \\
      8c O$^{(1)}$ & (0.112737, 0.612737, 0.887263) & (0.11325, 0.613253, 0.886747) \\
      24e O$^{(2)}$ & (0.873628, 0.884526, 0.380629) & (0.87316, 0.884637, 0.381276) \\ \hline
      \multicolumn{3}{c}{Bond lengths (\AA)}\\\hline
      Li-O	& 2.072	& \\
      Ga$_{t}$-O$^{(1)}$ & 1.867 & \\
      Ga$_{t}$-O$^{(2)}$ & 1.903 & \\
      Ga$_{o}$-O$^{(2)}$ & 1.986 & \\
      Ga$_{o}$-O$^{(2)}$ & 1.939 & \\
      Ga$_{o}$-O$^{(1)}$ & 2.039 & \\
          \end{tabular}
  \end{ruledtabular}
\end{table*}

\subsection{Group Theoretical Analysis}
Since there are 56 atoms in the cell there are 168 degrees of motion or vibrational modes. Of these, three are acoustic and correspond to a translation of the crystal at ${\bf q}=0$. Group theory allows one to classify the remaining modes according to their irreducible representations. The character table of the group $O$ is given in Table \ref{character}. For each type of Wyckoff position we can derive a reducible representation from the displacements in the three Cartesian directions with the characters shown in the second part of the table.  These can then be reduced to irreducible representations as shown in the last column. Since there are both tetrahedral Ga$_t$ and O$^{(1)}$ in 8c positions, the 168 modes break up into $\Gamma=6A_1+8A_2+14E+21T_1+20T_2$ modes. Of these one $T_1$ corresponds to the uniform translations and needs to be removed. Of the remaining 165 modes, only $T_1$ is infrared (IR) active because it corresponds to  the components of a vector. Thus we find 21 $T_1$ IR active  modes. These will show an LO-TO splitting.  Raman modes correspond to a symmetric second rank tensor and can thus belong to $A_1,E,T_2$  irreducible representations. So there are 
$6A_1+14E+20T_2$ Raman modes and the $8A_2$ modes are silent. 
The Raman modes of $T_2$ symmetry only have off-diagonal non-zero components $xy,yz,zx$ and are all three-fold degenerate. The $A_1$ Raman modes have a Raman tensor proportional to the unit matrix while the $E$ Raman modes are doubly degenerate and have only diagonal components as shown below.
\begin{table*}
\large
  \caption{Character Table of the group $O$ and reducible representations of the Wyckoff positions. \label{character}}
  \begin{ruledtabular}
    \begin{tabular}{ccccccc}
      Modes & E & 8\( C_3 \) & 6\( C_2' \) & 6\( C_4 \) & 3\( C_2 \) & Function \\
      \hline
      \( A_{1} \) & 1 & 1 & 1 & 1 & 1 & \( x^{2} + y^{2} + z^{2} \) \\
      \( A_{2} \) & 1 & 1 & -1 & -1 & 1 & -- \\
      \( E \) & 2 & -1 & 0 & 0 & 2 & \( x^{2} - y^{2},\; 2z^{2} - x^{2} - y^{2} \) \\
      \( T_{1} \) & 3 & 0 & -1 & 1 & -1 & \( x,\; y,\; z \) \\
      \( T_{2} \) & 3 & 0 & 1 & -1 & -1 & \( xy,\; yz,\; zx \) \\ \hline
      \(4b\)      & 12 & 0 & -2 & 0 & 0 &  \( A_2+E+2T_1+T_2\) \\
      \(8c\)      & 24 & 0 & 0 & 0 & 0 & \( A_1+A_2+2E+3T_1+3T_2\) \\
      \(12d\)     & 36 & 0 & -2 & 0 & 0 &  \(A_1+2A_2+3E+5T_1+4T_2\) \\
      \( 24e\)    & 72 & 0 & 0 & 0 & 0 & \(3A_1+3A_2+6E+9T_1+9T_2\)
    \end{tabular}
  \end{ruledtabular}
\end{table*}

\[
\begin{array}{cc}
\begin{aligned}
A_{1} &= 
\begin{pmatrix}
a & . & . \\
. & a & . \\
. & . & a
\end{pmatrix}
\end{aligned}
&
\begin{aligned}
E &= 
\begin{pmatrix}
b & . & . \\
. & b & . \\
. & . & -2b
\end{pmatrix}
\end{aligned}
\\
\\[1.5em]
\begin{aligned}
E &= 
\begin{pmatrix}
-\sqrt{3}b & . & . \\
. & \sqrt{3}b & . \\
. & . & .
\end{pmatrix}
\end{aligned}
&
\begin{aligned}
T_{2} &= 
\begin{pmatrix}
. & . & b \\
. & . & . \\
b & . & .
\end{pmatrix}
\end{aligned}
\\
\\[1.5em]
\begin{aligned}
T_{2} &= 
\begin{pmatrix}
. & b & . \\
b & . & . \\
. & . & .
\end{pmatrix}
\end{aligned}
&
\begin{aligned}
T_{2} &= 
\begin{pmatrix}
. & . & . \\
. & . & b \\
. & b & .
\end{pmatrix}
\end{aligned}
\\
\end{array}
\]
\subsection{Electronic band structure}\label{secelectron}
\begin{figure}[h]
\centering
\includegraphics[width=\columnwidth]{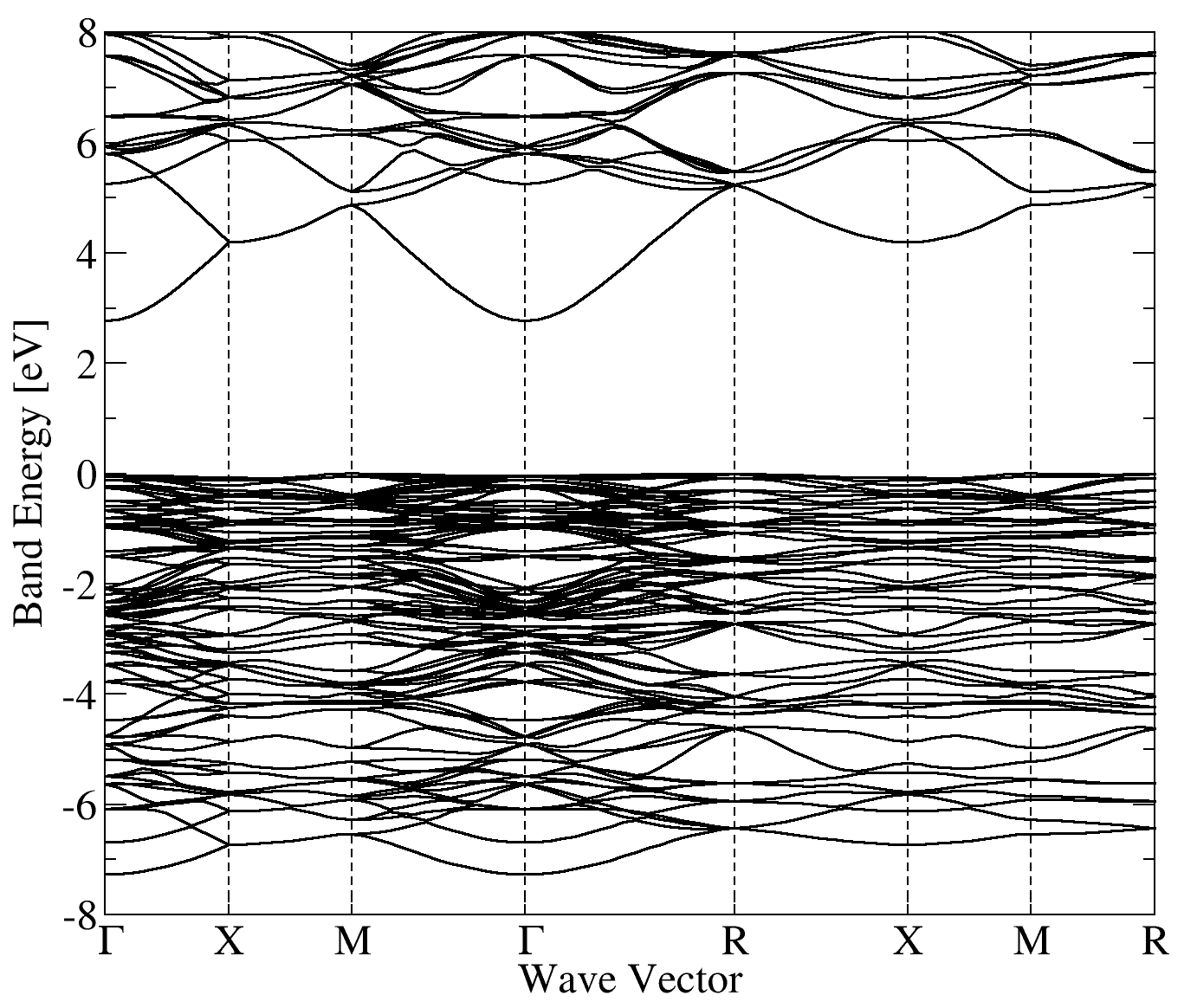}
\caption{Electronic band structure of LiGa$_{5}$O$_{8}$ along high-symmetry path, $X=(\pi/a,0,0), M=(\pi/a,\pi/a,0),R=(\pi/a,\pi/a,\pi/a),\Gamma=(0,0,0)$}
\label{figelbands}
\end{figure}
\begin{figure}[h]
\centering
\includegraphics[width=\columnwidth]{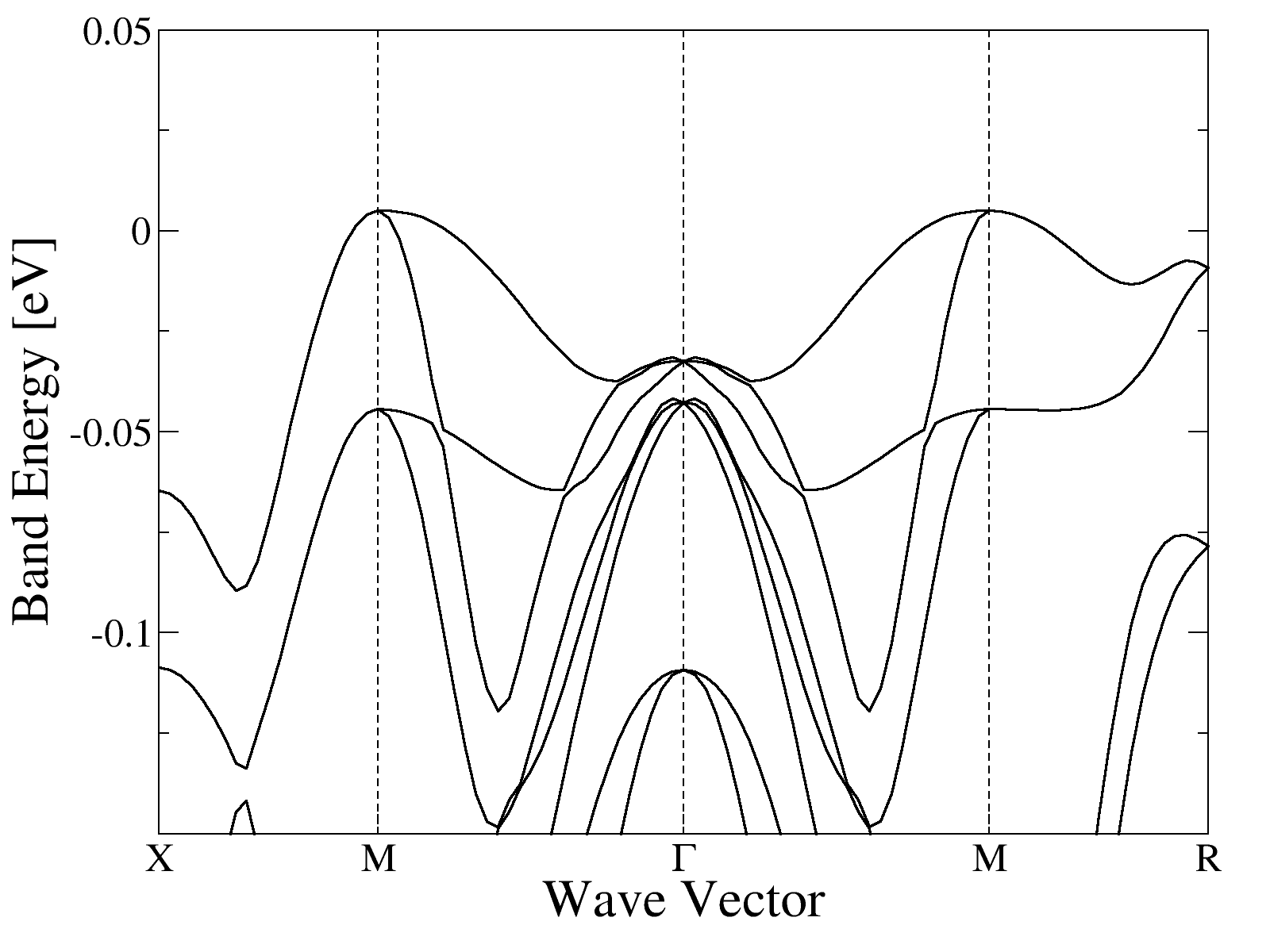}
\caption{Band structure near the valence band maximum of LiGa$_{5}$O$_{8}$ at $M=(\pi/a,\pi/a,0)$}
\label{figvbm}
\end{figure}

While the electronic band structure has been already studied with superior methods such as the QS$GW$ method in \cite{Lambrecht2024} we here check the band structure obtained in the PBEsol functional. The electronic band structure is shown in Fig. \ref{figelbands} and shows a slightly smaller indirect gap. The direct 2.86eV gap at $\Gamma$ and indirect gap 2.76eV  found here are consistent with the GGA gaps given in that paper which are respectively 2.58 eV and 2.48 eV. That paper also used the PBEsol functional but a different all-electron method  and  the full-potential linearize muffin-tin orbital method. It shows that the convergence of our present PAW calculation is adequate. 

In Fig.\ref{figvbm} we show a zoom in of the band structure near the valence band maximum (VBM). We can see that the VBM occurs at $M$ and the band is doubly degenerate along $MX$ and $MR$ on  the Brillouin zone surface but splits along $M\Gamma$ into a heavy hole and light hole band. These two bands near $M$ can be described by an effective $2\times2$ Hamiltonian  of the form 
\begin{equation}
    H=[A(q_z^2+q_y^2)+Bq_z^2]\bm{\delta}+C\bm{\sigma}_xq_xq_y,
\end{equation} 
with ${\bf q}$ the deviation from point $M$, $\bm{\delta}$  the $2\times2$ unit matrix and $\bm{\sigma}_x$ the Pauli matrix. The inverse mass parameters $A,B,C$ are related to $A=1/m_{MX}$, $B=1/m_{MR}$ which are the components of the effective mass tensor of the degenerate band in $x$ and $z$ direction, while $2A\pm C=2/m_{l(h)h}$ give the inverse heavy and light hole masses along $M\Gamma$. 
The conduction band minimum (CBM) at $\Gamma$ has an isotropic mass tensor $m_c$ because of the cubic symmetry. These masses can be determined simply by fitting a $q^2$ behavior to the dispersion near the band extrema and lead to the values given in Table \ref{tabmass}.
We here  give the masses both extracted from  the current PBEsol band structure results and from  the all-electron PBESol and QS$GW$ results from \cite{Lambrecht2024}. Note that the heavy massses suffer from larger uncertainty due to the small changes of the band over the region where parabolic dispersion holds and are sensitive to the range of k used in the fit. However, within each set  of results, the relation between the effective masses along $M\Gamma$  and $MX$  predicted by the effective Hamiltonian hold fairly well.  PBEsol is seen to overestimate the masses relative to the more accurate QS$GW$.
For the conduction band mass the density functional perturbation theory approach as implemented in Janssen 
{\sl et. al.}\cite{Janssen2016} gives the same result as the direct fit and good agreement between the different approaches is obtained.  For the VBM the double degeneracy complicates the calculation of the effective mass tensor. For $M$ in the $xy$-plane, we find a high effective mass along the $z$-direction  and for the top band along the [110] direction but there are 12 equivalent $M$ points at $(\pm\pi/a,\pm\pi/a,0), (\pm\pi/a,0,\pm\pi/a)$, $(0,\pm\pi/a,\pm\pi/a)$, and hence for every $M$ there is  some direction with heavy mass. These heavy masses found here are similar to those in Ga$_2$O$_3$\cite{Ponce,Mock} and indicate that hole mobilities may be low and self-trapped polarons are likely to occur.

\begin{table}
 \caption{Effective masses (in units of the free electron mass).\label{tabmass}} 
 \begin{ruledtabular}
 \begin{tabular}{lccc}
 &this work&PBEsol\cite{Lambrecht2024}& QS$GW$\cite{Lambrecht2024}\\ \hline
 $m_c$ & 0.27 & 0.26 & 0.26\\
 $m_{hh}$ & 10.28 & 7.72  & 4.02\\
 $m_{lh}$ &0.77 & 0.66 & 0.48 \\
 $m_{MX}$&  1.47 & 1.21 & 0.87 \\
 $m_{MR}$ &6.88 & 5.63 & 3.34\\ \hline
 $A$      & 0.70 &0.82 & 1.17 \\
 $B$      & 0.15 & 0.18  & 0.30\\
 $C$      & 1.21 & 1.39  &1.83 \\
 \end{tabular}
 \end{ruledtabular}
\end{table}

\subsection{Phonon dispersions and density of states}\label{secphondos}
\begin{figure}[h]
  \centering
  \includegraphics[width=\columnwidth]{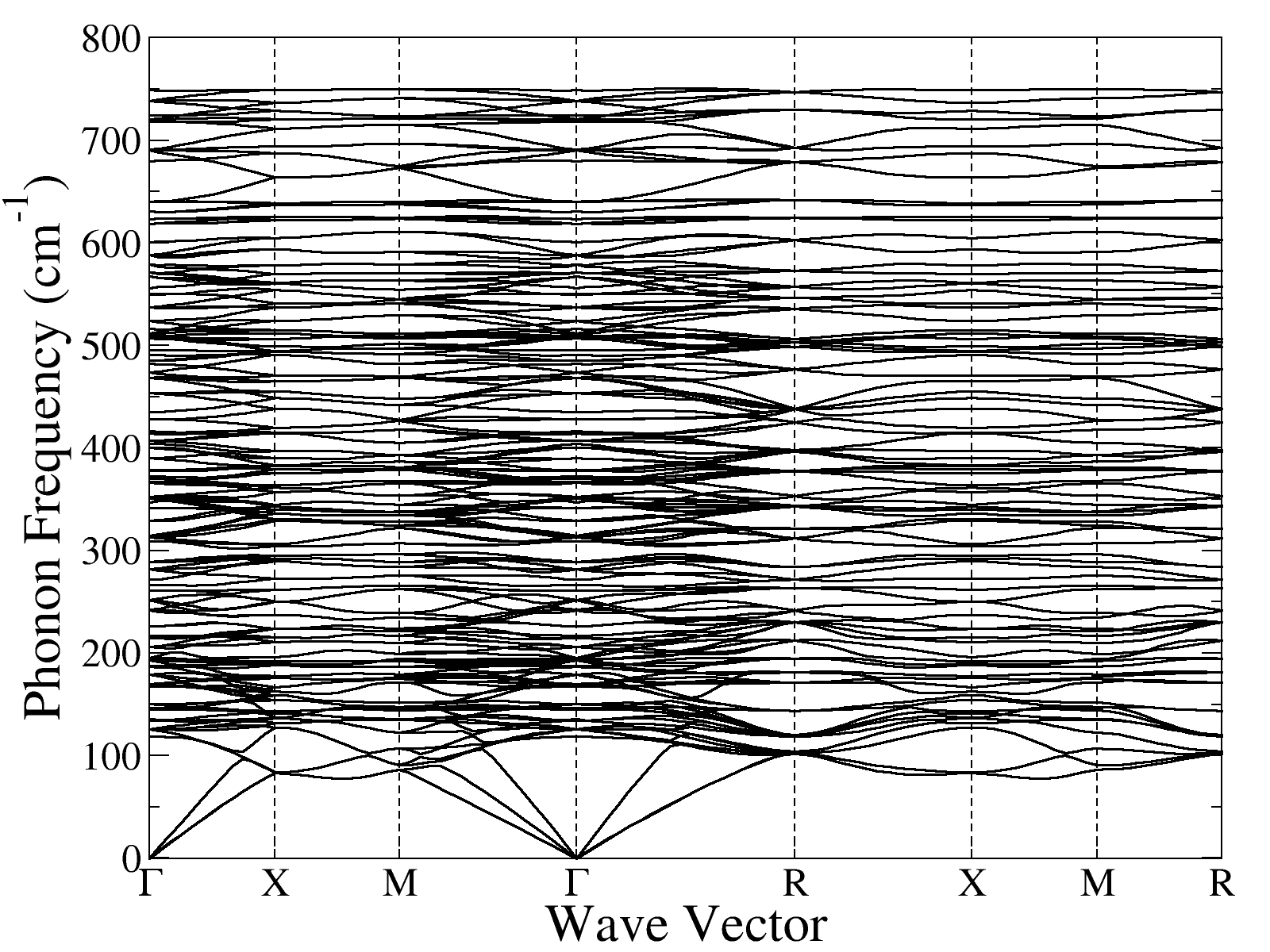}
  \caption{Phonon bandstructure of LiGa$_{5}$O$_{8}$ along high-symmetry path.}
  \label{phononband}
\end{figure}

The phonon band structure  is shown in Fig. \ref{phononband}. First, we note that there are no imaginary frequency modes, which verifies the mechanical stability of the structure. We can recognize the three lowest acoustic modes followed by 165 modes. The dips near $\Gamma$ reflect the LO-TO splitting of the IR active $T_1$ modes. The values at $\Gamma$ itself represent the TO frequency while the limit ${\bf q}\rightarrow 0$ represent the LO modes. 

\begin{figure}[h]
  \centering
  \includegraphics[width=\columnwidth]{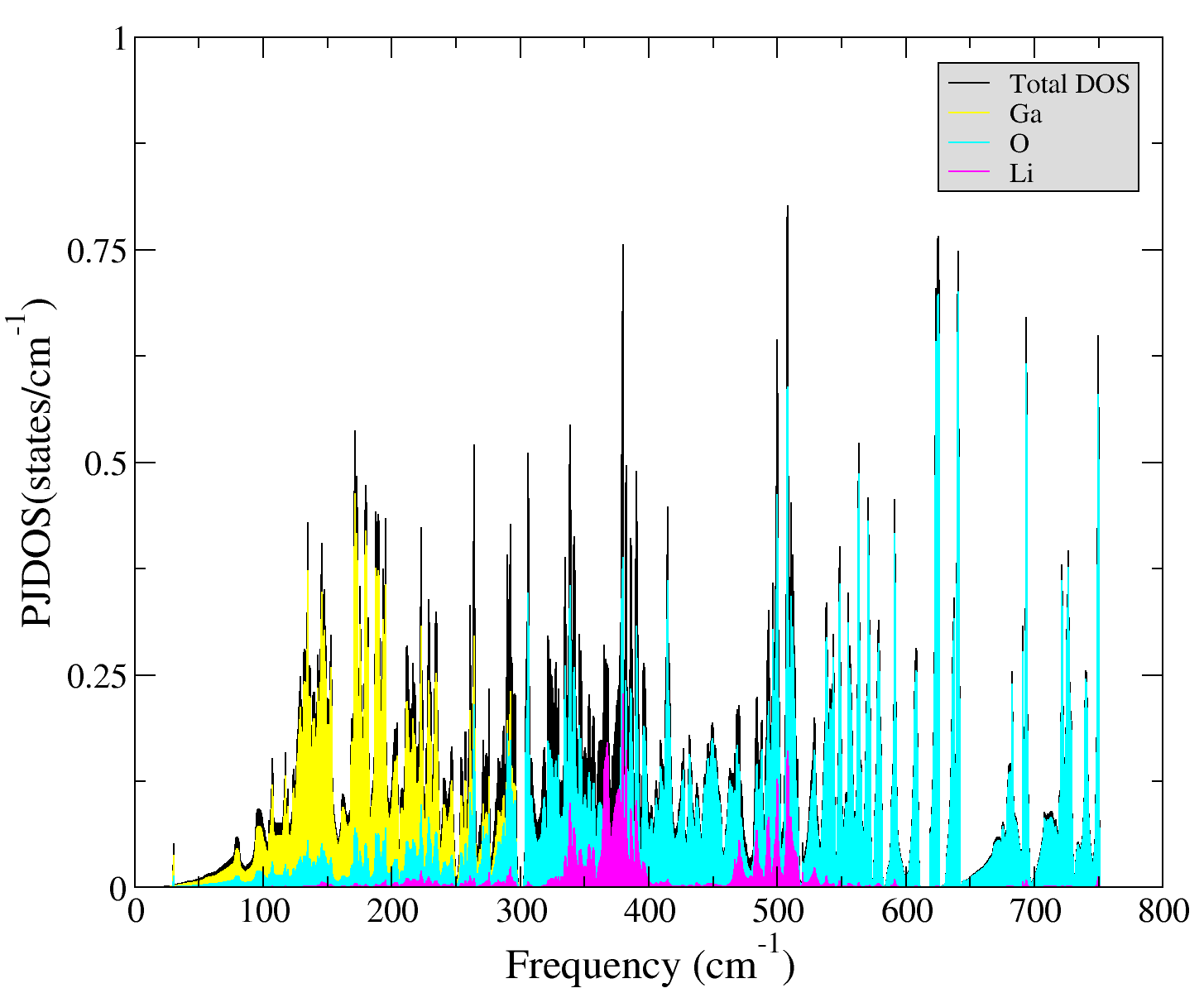}
  \caption{Phonon projected density of states of LiGa$_{5}$O$_{8}$.}
  \label{phonondos}
\end{figure}

Next, we show the Phonon Density of states (DOS) in Fig.\ref{phonondos}. 
It shows the usual quadratic onset in the region of acoustic modes. The decomposition in atomic components shows that the highest modes have primarily O character. Even though Li atoms  have smaller mass, their main contributions appear between 300-400 and near 500 cm$^{-1}$. This indicates that the Li-O bonds have smaller force constants than some of the Ga-O ones.
This is consistent with the bond-lengths in Table \ref{lattice}. The lower energy range up to 250 cm$^{-1}$ shows primarily Ga contribution as expected because Ga is the heaviest atom in the system. A phonon DOS like spectrum might be expected in Raman spectra of samples with significant disorder breaking the crystal momentum conservation, while highly crystalline samples would show only zone centrum modes. 
Furthermore a reduction of phonon DOS in the regions 300-400 cm$^{-1}$ and 500 cm$^{-1}$ in observed spectra might indicate the presence of Li vacancies while an enhancement in the low energy region might indicate  Ga rich  stoichiometry or a reduction in this low energy region, Ga vacancies. A reduction of the highest region of the spectra might be indicative of oxygen vacancies. Thus, the Raman spectra might provide indirect information on the defects present in the material. 

\subsection{Phonons at the zone center}\label{secphoncenter}

The numerical values of the phonons at $\Gamma$ are 
 presented in Table~\ref{phonon_modes}  and sorted according to their symmetry character. At the zone center (\textbf{q}=0), due to the long-range interaction between the macroscopic electric field and the Born effective charge $Z^{*}_{\kappa,\beta\alpha}$ LO-TO splitting is present for the $T_1$ modes.  We can see that  for the low frequency range the splitting is not so prominent, but in the high frequency ($>397.7$ cm$^{-1}$) regime the splitting is significant. This reflects the stronger dipolar character of these modes, which correspond to the various bond stretch modes. 
\\
\begin{table*}
\large
  \caption{LiGa$_5$O$_8$ Phonon Modes Symmetry Labeling (in $cm^{-1}$ unit) \label{phonon_modes}}
  \begin{ruledtabular}
    \begin{tabular}{|c|c|c|cc|cc|cc|}
 $A_1$ & $A_2$ & $E$ &  \multicolumn{2}{c|}{$T_{1,\textit{TO}}$} &  \multicolumn{2}{c|}{$T_{1,\textit{LO}}$} &  \multicolumn{2}{c|}{$T_{2}$} \\
 \hline
123.3 & 150.2 & 129.8 & 138.2 & 467.5 & 138.2	&   483.3 & 129.3 & 517.4		\\
258.9 & 193.5 & 148.1 & 172.8 & 491.9 & 172.8	&   492.4 & 150.6 & 544.7		\\
403.6 & 273.9 & 169.1 & 189.5 & 498.7 & 189.8	&   529.6 & 177.3 & 570.1		\\
484.5 & 360.6 & 225.8 & 193.3 & 549.3 & 193.6	&   571.7 & 194.2 & 614.5		\\
525.6 & 469.5 & 260.9 & 215.2 & 587.5 & 216.9	&   655.6 & 204.1 & 666.4		\\
715.1 & 495.7 & 306.5 & 239.2 & 665.3 & 239.9	&   693.4 & 246.2 & 694.6		\\
      & 520.3 & 323.4 & 276.2 & 714.8 & 276.2	&   724.1 & 288.7 &			\\
      & 698.8 & 352.9 & 303.6 &       & 303.7	&	  & 306.0 &			\\
      &       & 379.7 & 320.6 &       & 331.0	&	  & 327.6 &			\\
      &       & 407.1 & 332.6 &	      & 359.0	&	  & 362.2 &			\\
      &       & 499.8 & 368.5 &	      & 370.8	&	  & 394.0 &			\\
      &       & 546.8 & 390.9 &	      & 394.6	&	  & 438.0 &			\\
      &       & 593.0 & 397.7 &	      & 424.4	&	  & 455.8 &			\\
      &       & 603.7 & 458.3 &	      & 462.2	&	  & 493.7 &			\\
    \end{tabular}
  \end{ruledtabular}
\end{table*}
\\
\subsection{IR Spectra and Dielectric Properties}\label{secir}
The infrared optical response is fully described by the macroscopic dielectric permittivity tensor, $\varepsilon(\omega)$. 
It is obtained from the oscillator strengths $S_{m,\alpha\alpha}$, as 
\begin{equation}
 \varepsilon_{\alpha\alpha}(\omega)=\varepsilon_{\alpha\alpha}^{\infty}+\frac{4\pi}{\Omega_0}\frac{\sum_mS_{m,\alpha\alpha}}{\omega_m^2-\omega^2-i\Gamma_n},   
\end{equation}
where $\varepsilon_{\alpha\alpha}^\infty$ is the purely electronic screening  contribution. The oscillator strengths in turn  depend on the Born effective charges and eigenmode eigen vectors,
\begin{equation} 
S_{m,\alpha\alpha}=|\sum_{\kappa\alpha'}Z^{*}_{\kappa,\alpha\alpha'}U^{*}_{m\textbf{q=0}}(\kappa\alpha')|^2
\end{equation}
Here, $\kappa$ indicates the atom number in the unitcell and $\alpha,\alpha'$ the Cartesian directions.
For reference, the diagonal elements of the Born effective charge tensors  are provided in Table \ref{tabborn} and the oscillator strenths for the $T_1$ modes in Table~\ref{tabosc}. We note that in a cubic crystal the oscillator strength matrix has only diagonal components and they are all equal so there is only one component per mode $m$. 
The Born effective charges are defined by 
\begin{equation}
  Z^\ast_{\kappa,\beta\alpha}=\Omega_0\frac{\partial\mathcal{P}_{mac,\beta}}{\partial\tau_{\kappa\alpha}(\textbf{q=0})}=\frac{\partial F_{\kappa,\alpha}}{\partial \varepsilon_\beta} 
\end{equation}
They do have small off-diagonal elements which add up to zero when summed over all equivalent atoms of a given Wyckoff position type. The Born charges of Li, Ga and O are seen to be close to their nominal values of +1, +3 and $-2$ respectively but not all diagonal components are equal. This is because the Born effective charges reflect site symmetry of the Wyckoff position rather than the global symmetry. The charge neutrality  rule $\sum_\kappa Z^*_{\kappa\alpha\beta}=0$ for each pair of Cartesian components is enforced by renormalizing the initial calculated ones.  Of the 12 octahedral Ga, 4 have the $Z_{xx}$ value listed in the table and 8 have the $Z_{yy}=Z_{zz}$ value listed in the table as their $Z_{xx}$. Of the 24 O$^{(2)}$, 8 have have the $Z_{xx}$, 8 the $Z_{yy}$ and 8 the $Z_{zz}$ value listed. Thus $4Z_{xx}^{\rm Li}+8Z_{xx}^{{\rm Ga}_t}+4Z_{xx}^{{\rm Ga}_o}+8Z_{yy}^{{\rm Ga}_o}+8 Z_{xx}^{{\rm O}1}+8 Z_{xx}^{{\rm O}2}+8Z_{yy}^{{\rm O}2}+8Z_{zz}^{{\rm O}2}=0$. We also provide the calculated high frequency and static dielectric  constants in Table\ref{tabdiel}.

\begin{figure}[h]
  \centering
  \includegraphics[width=\columnwidth]{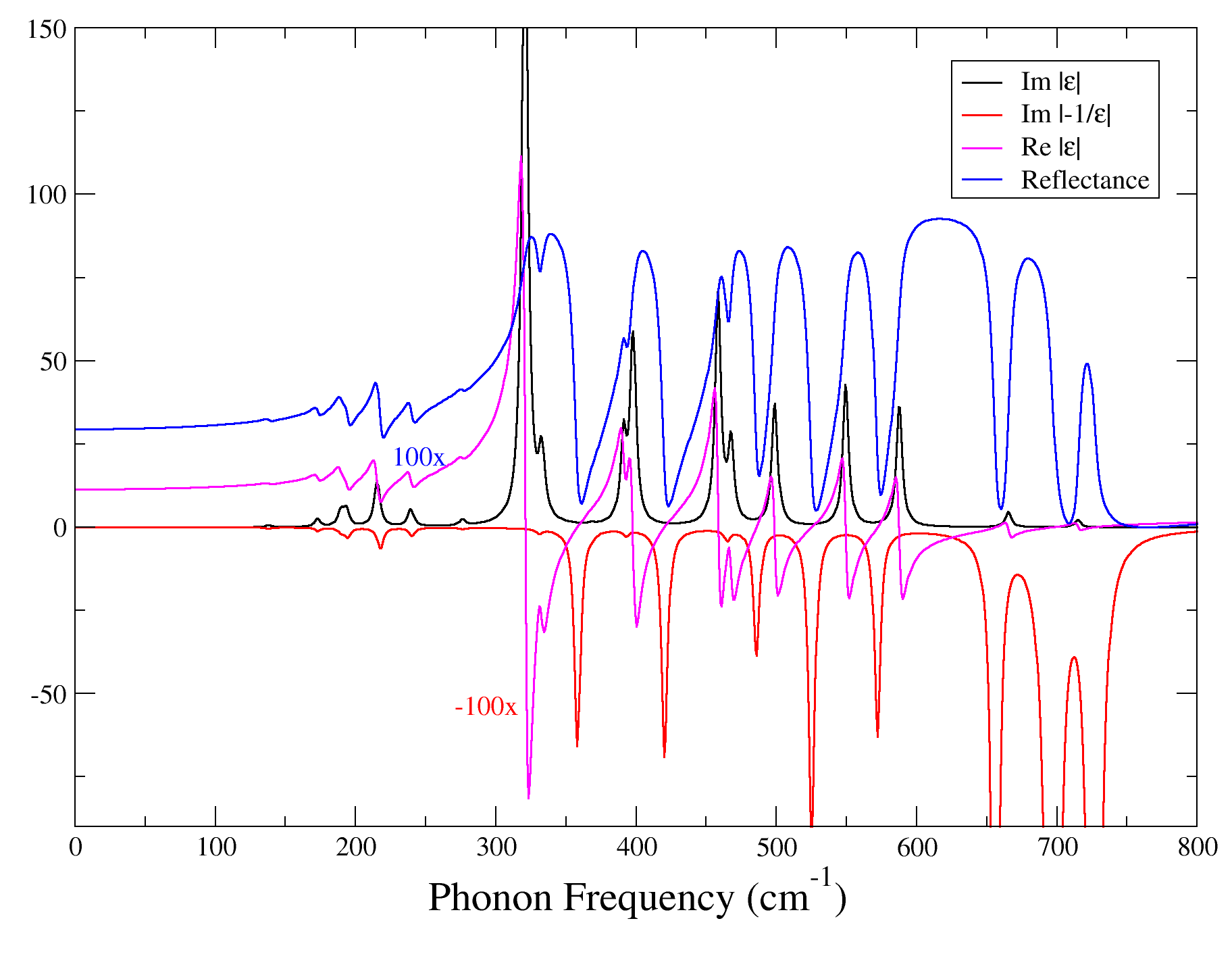}
  \caption{\large Dielectric properties and IR spectra of LiGa$_{5}$O$_{8}$.}
  \label{IR}
\end{figure}

\begin{figure}
\includegraphics[width=8cm]{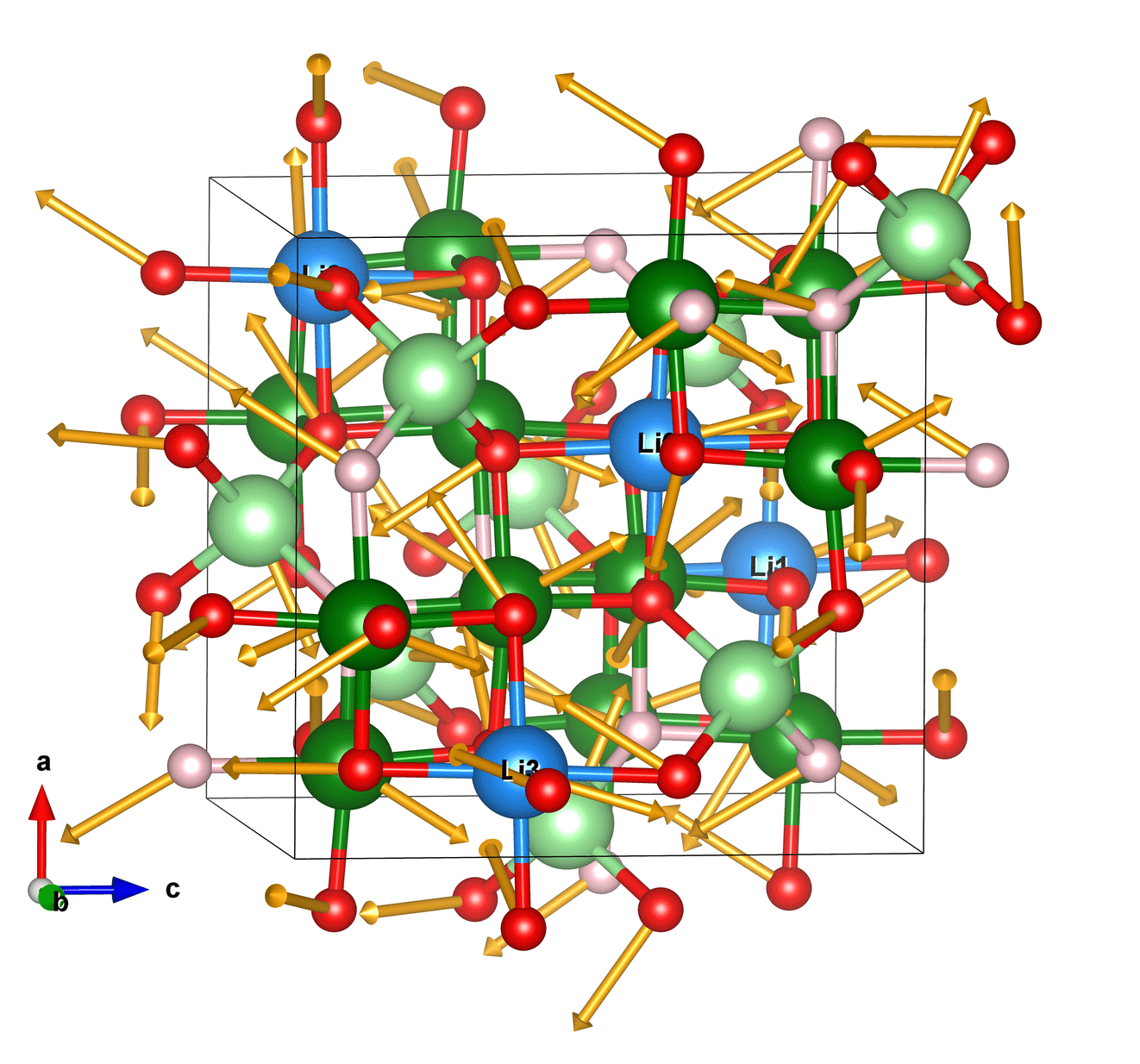}
\caption{Eigen displacements of mode $T_1^9$  at 320.6 cm$^{-1}$}\label{figmodet1-9}
\end{figure}

The poles of the dielectric function, which correspond to peaks in its imaginary part $\varepsilon_2(\omega)$ correspond to the TO modes, while the zeros of the real part $\varepsilon_1(\omega)$ correspond to the LO modes. The latter also show up as peaks of the loss function $-\Im{[\varepsilon^{-1}(\omega)]}$. From these, one can extract the absorption coefficient $\alpha(\omega)=\omega\varepsilon_2(\omega)/n(\omega)c$ with $\tilde n=n+i\kappa=\sqrt{\varepsilon}$ the complex index of refraction and the normal incidence reflectance $R=|\frac{\tilde{n}-1}{\tilde{n}+1}|$ 
 Using a fixed broadening  parameter $\Gamma_n$ of 5 cm$^{-1}$ for all modes, we obtain the dielectric function and related properties, displayed in Fig.~\ref{IR}. This is a typical value used to obtain a smooth spectrum. The actual phonon linewidth may vary from mode to mode and has contributions from electron-phonon and three- and four-phonon scattering anharmonic terms, and are here not calculated \cite{Han2022,WULi2014,Han2022-4ph}.
When experimental spectra become available, one could adjust the linewidth to obtain a better fit but in the absence of such information, the single value used here is sufficient to provide a first approximation to the spectral shape. 
 Because some modes, notably modes 8(61-63), 11 (86-88), 16(117-119),  have very small oscillator strength (and small TO-LO) splitting, they are not clearly visible and hence we see only 18 instead of the predicted 21 peaks.  Note that the strongest mode is mode $T_1^9$ which corresponds to modes 69-71 when ordering all modes irrespective of symmetry according to energy as given in  Table \ref{tabosc}.  
 As an example, we show the eigendisplacement pattern for mode $T_1^9$ in Fig. \ref{figmodet1-9}. We can see that most of the cations have an upward motion while the O atoms have a downward motion. Thus, there is a significant dipole character to the overall motions consistent with a large IR active oscillator strength. There are both Li-O and Ga-O bond stretches involved in this mode and they are in phase for equivalent atoms.
 
\begin{table}
\large
  \caption{Born Effective Charges. \label{tabborn}}
  \begin{ruledtabular}
    \begin{tabular}{cccc}
Atom    &       $Z_{xx}$     &      $Z_{yy}$     &      $Z_{zz}$     \\ \hline
Li      &       1.081846 &      1.081846 &      1.081846 \\
Ga$_t$    &       2.964809 &      2.964809 &      2.964809 \\
Ga$_o$   &       3.202848 &      3.199848 &      3.199848 \\
O$^{(1)}$     &      -2.142370 &     -2.142370 &     -2.142370 \\
O$^{(2)}$    &      -2.200546 &     -2.287746 &     -1.676342 \\
    \end{tabular}

  \end{ruledtabular}
\end{table}

\begin{table}
\large
  \caption{Oscillator Strenghts \label{tabosc}}
  \begin{ruledtabular}
    \begin{tabular}{cc}
Mode Number	& $S_{xx}$ = $S_{yy}$ = $S_{zz}$	\\
\hline
10 - 12	  & $2.4051\times10^{-6}$	\\
21 - 23	  & $1.2157\times10^{-5}$	\\
27 - 29	  & $2.6133\times10^{-5}$	\\
33 - 35	  & $2.8206\times10^{-5}$	\\
40 - 42	  & $8.7932\times10^{-5}$	\\
45 - 47	  & $3.8607\times10^{-5}$	\\
55 - 57	  & $1.5080\times10^{-5}$	\\
61 - 63	  & $9.0217\times10^{-7}$	\\
69 - 71	  & $1.9435\times10^{-3}$	\\
77 - 79	  & $1.9880\times10^{-4}$	\\
86 - 88	  & $3.9548\times10^{-6}$	\\
91 - 93	  & $3.0503\times10^{-4}$	\\
97 - 99	  & $6.9824\times10^{-4}$	\\
109 - 111 & $9.9194\times10^{-4}$	\\
112 - 114 & $3.4856\times10^{-4}$	\\
117 - 119 & $2.5917\times10^{-8}$	\\
126 - 128 & $5.7461\times10^{-4}$	\\
139 - 141 & $7.3431\times10^{-4}$	\\
145 - 147 & $6.7895\times10^{-4}$	\\
155 - 157 & $9.5026\times10^{-5}$	\\
166 - 168 & $4.9986\times10^{-5}$	\\
\end{tabular}

\end{ruledtabular}
\end{table}

\begin{table}
\large
  \caption{Dielectric constants} \label{tabdiel}
  \begin{ruledtabular}
    \begin{tabular}{cc}
$\varepsilon_\infty$ & $\varepsilon_0$ \\ \hline
4.107 & 10.959\\
\end{tabular}
\end{ruledtabular}
\end{table}

\subsection{Raman spectra}\label{secraman}
\begin{figure*}
\centering
\begin{minipage}{.5\textwidth}
  \centering  \includegraphics[width=0.9\linewidth]{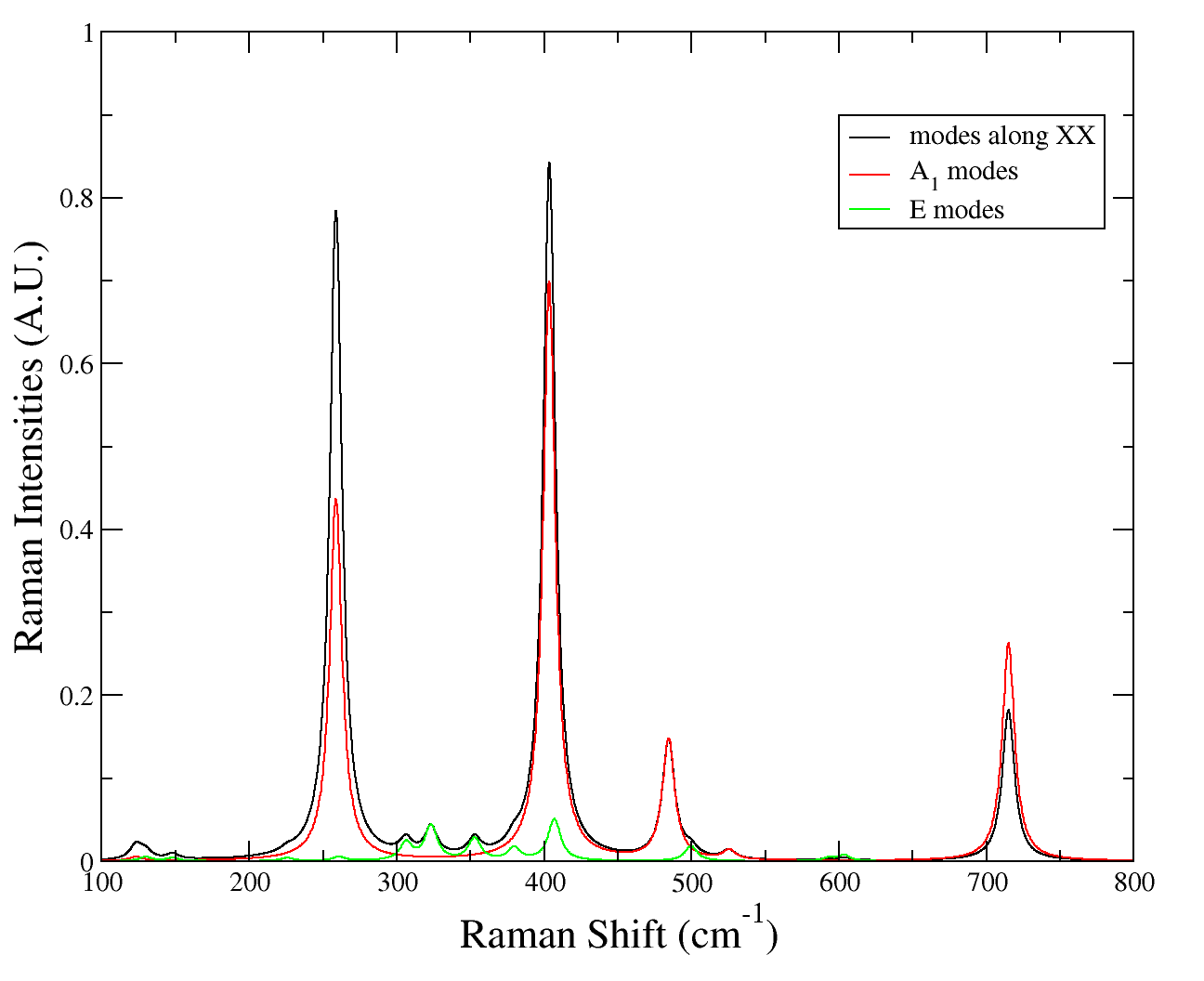}
  \caption{Simulated Raman spectra of LiGa$_{5}$O$_{8}$ for $A_1$ and E modes.}
  \label{figramanpar}
\end{minipage}%
\begin{minipage}{.5\textwidth}
  \centering  \includegraphics[width=0.9\linewidth]{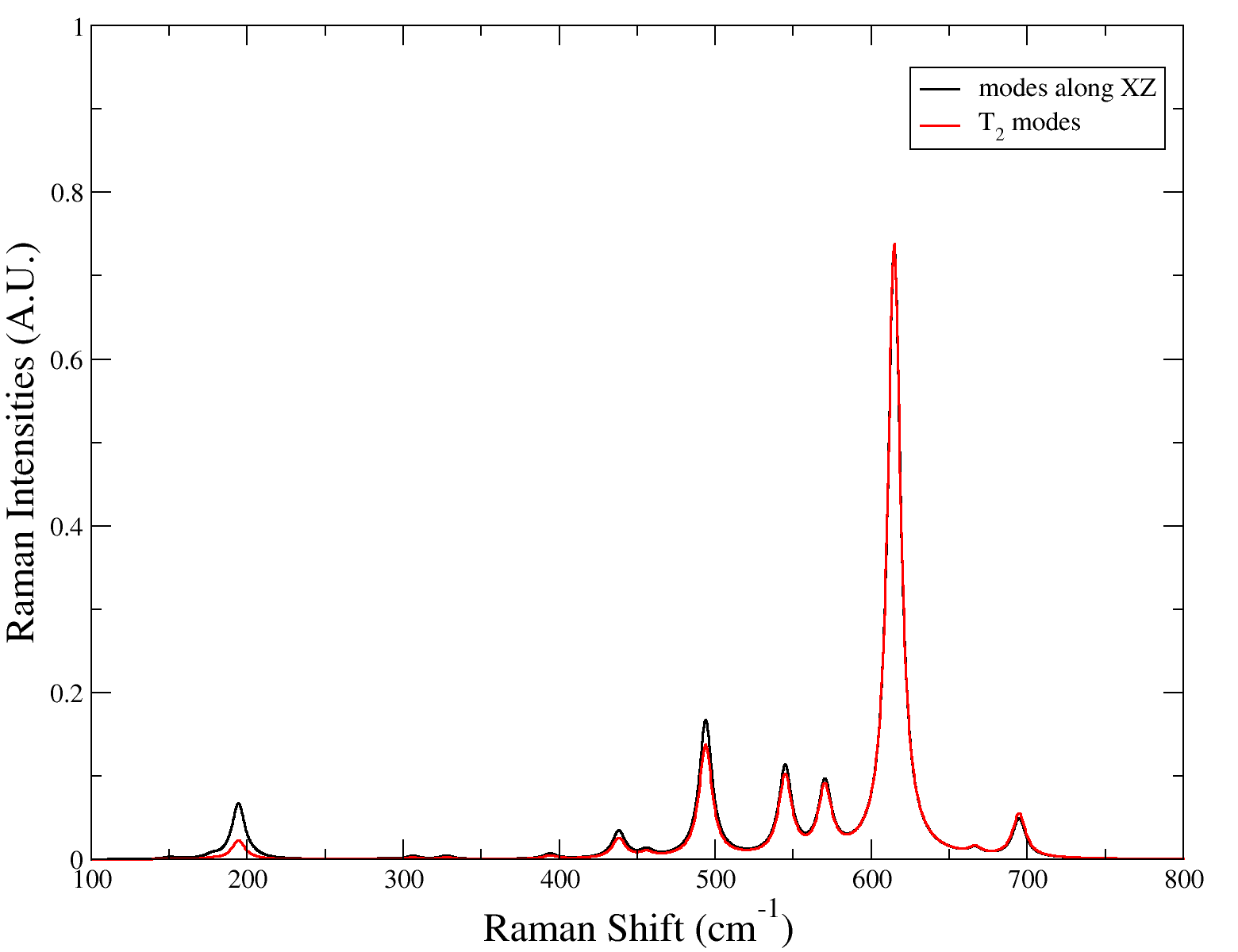}
  \caption{Simulaed Raman spectra of LiGa$_{5}$O$_{8}$ for $T_2$ modes.}
  \label{figramant2}
\end{minipage}
\end{figure*}

The Raman simulated spectra are obtained by broadening the modes into a Gaussian peak with the same uniform linewidth of 5 cm$^{-1}$ as used for the IR spectra. Again, this is just a typical value and may require adjustment for a better fit to experiment. 
In Fig.\ref{figramanpar} we show the Raman modes of $A_1$ and $E$ symmetry. Note that the $E$- modes have significantly smaller intensity. Both of these are active in a scattering geometry with parallel incident and scattered light polarizations.
We show the spectra for $T_2$ modes separately  in Fig.\ref{figramant2} because these are active for perpendicular incident and 
\begin{figure}
\includegraphics[width=8cm]{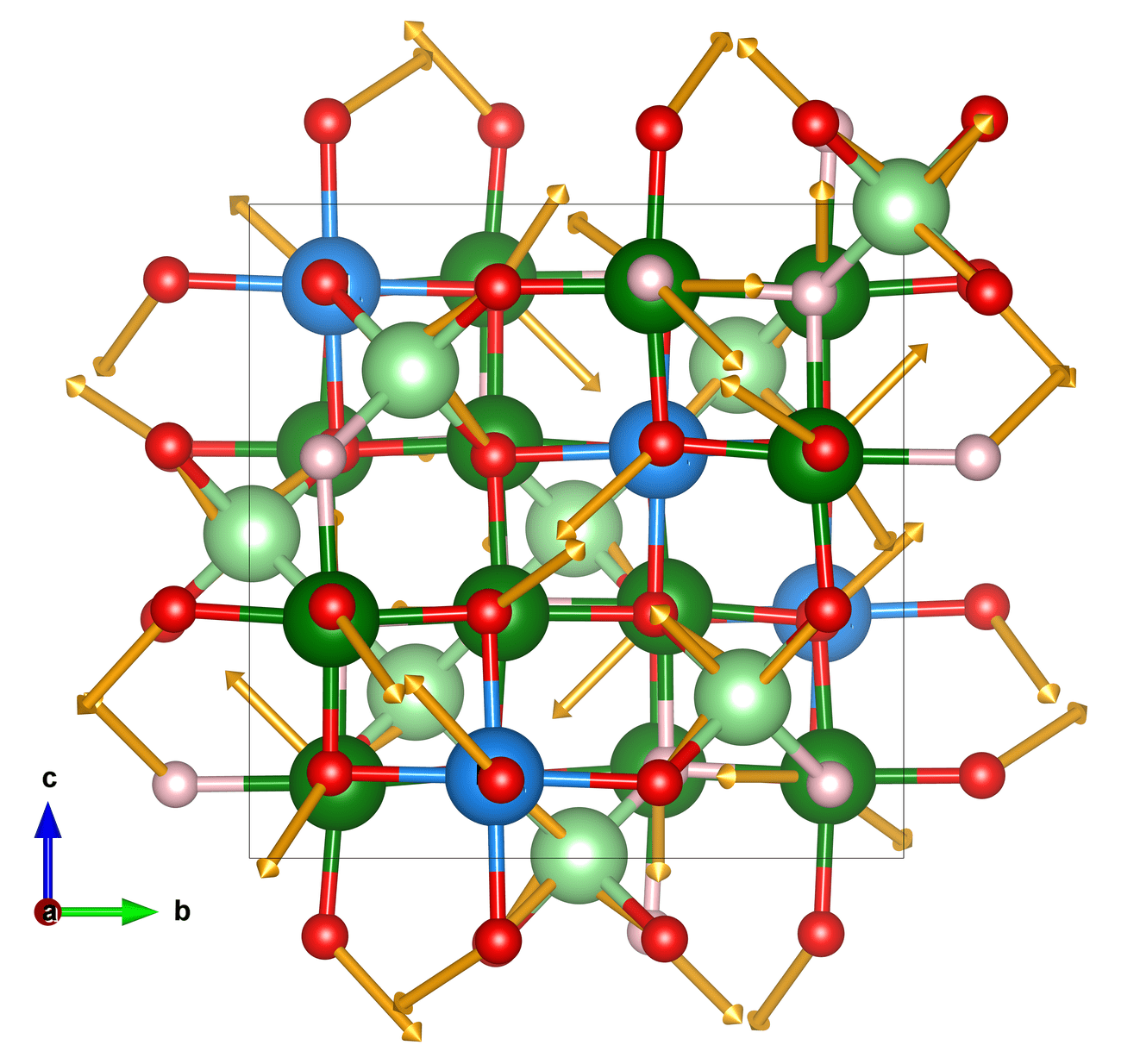}
\caption{Eigendisplacement pattern for mode $A_1^6$} at 715.1 cm$^{-1}$\label{figa1}
\end{figure}

\begin{figure*}[ht]
  \centering
\includegraphics[width=40em]{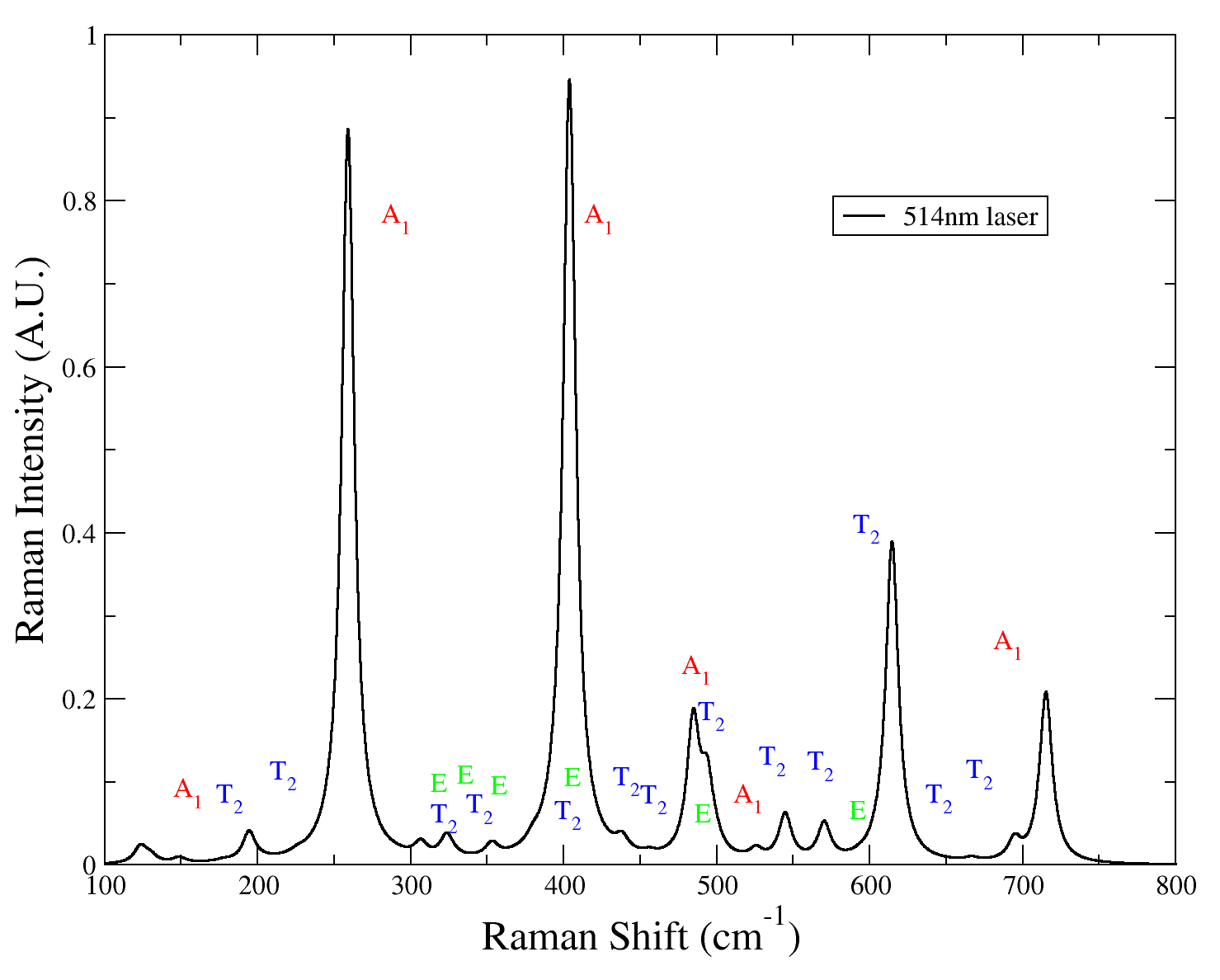}
  \caption{\large Simulated polycrystalline angle averaged Raman spectrum of LiGa$_{5}$O$_{8}$ for 514nm laser incident wavelength.}
  \label{figramanpoly}
\end{figure*}

the Raman spectrum  averaged over directions which is relevant for a polycrystalline sample  is shown in Fig.\ref{figramanpoly}.

In Fig.\ref{figa1} we illustrate the eigen displacements of the highest frequency  $A_1^6$ mode. We can see several bond stretches but now they are out of phase between different equivalent ones so that the net oscillator strength is zero. This mode preserves the symmetry of the crystal.  Note that in this mode the Li atoms do not participate. 

We finally note that these represent first-order Raman scattering  and assume crystal momentum conservation so that only modes at the zone center  contribute. Often in samples with significant defect induced disorder, the momentum  conservation is relaxed and one then finds a disorder induced Raman scattering which can have contributions from the whole Brillouin zone and therefore is closer to  the integrated Phonon DOS which was presented in Sec.\ref{secphondos}. 

\section{CONCLUSION}
A first-principles study was presented of the phonon related properties of LiGa$_5$O$_8$. 
The PAW calculations in the PBEsol exchange correlation functional provide good agreement with experimental data on the structural parameters and a band gap consistent with previous calculations. The absence of imaginary modes confirms the mechanical stability of the spinel structure. The phonons at the zone center were  calculated and symmetry labeled. There are 21 infrared active $T_1$ modes, which could be identified in the calculated IR spectra and for which the LO-TO splitting was reported. There are 20 $T_2$ Raman active modes for perpendicular input and output polarizations, 14 $E$ modes and 6 $A_1$ modes which are Raman active under parallel polarizations and 8 silent $A_2$ modes. The dielectric IR and Raman spectra as well as supporting quantities such as the oscillator strengths, high-frequency and static dielectric constants, and the Born effective charges.  

{\bf Data Availability} The digital data related to the figures are p rovided in \url{https://github.com/Electronic-Structure-Group/Phonons-LiGa5O8}.

\acknowledgements{This work was supported by the US. Air Force Research Office (AFOSR)
  under grant No. FA9550-22-1-0201. It made use of the High Performance Computing Resource in the Core Facility for
  Advanced Research Computing at Case Western Reserve University}
\FloatBarrier
\bibliography{liga5o8,dft}

\end{document}